\documentclass[a4paper,12pt]{article}
\usepackage[T1,T2A]{fontenc}
\usepackage[utf8]{inputenc}
\usepackage[english]{babel}

\usepackage[a4paper]{geometry}
\usepackage{amsmath}

\usepackage{graphicx}
\usepackage[english]{babel}
\usepackage{authblk}

\title{\uppercase{On fast charged particles scattering on periodic atomic planes}}

\usepackage{lipsum}

\usepackage{multicol}
\usepackage{subcaption}
\usepackage{amssymb}
\usepackage{placeins}

\author[1]{V.D.~Omelchenko \thanks{koriukina@kipt.kharkov.ua}}
\affil[1]{National Science Center ``Kharkiv Institute of Physics and Technology'', Kharkiv, Ukraine}

\begin{document}
\maketitle

\begin{abstract}
{In this paper, the approach for considering fast charged particles scattering on targets of complex structure, which contain some isolated substructures, is introduced. Based on this approach, the differential cross section for scattering on the set of parallel planes with uniformly distributed atoms in each plane is obtained. It is shown that without account of thermal vibrations, the differential scattering cross section may be expressed analytically which simplifies the calculations.  
\par}

\end{abstract}

\section*{{Introduction}}
The study of fast charged particles is an important part of research in high-energy physics. This problem, considering the representation of free particles as plane electromagnetic waves, is related to the diffraction of electromagnetic waves in targets. In Bragg's work \cite{Bragg12_1}, the connection between the crystal structure and the characteristics of electromagnetic wave diffraction was demonstrated. The Laue-Bragg structure factor determines the differential scattering cross section in the Born approximation \cite{AIA96}, as shown in Ter-Mikaelian’s work \cite{MLT72}. Also, it is worth reminding of the scattering type called rainbow scattering. From the quasiclassical point of view, it occurs when same transffered momenta or deflaction angles can be obtained from scattering with different impact parameters. The rainbow scattering is also sensitive to the structure of the target \cite{Petrovic04}. A quantum description of the rainbow scattering \cite{FominR} is of particular interest, since in this case interference between different branches of the scattered particle wave function can be observed \cite{AIA96, Petrovic13}.

The description of high-energy charged particles scattering is based on quantum electrodynamics \cite{AIA96, AIA65}. Due to the complexity of the potential of the targets on which the particles are scattered, it makes sense to consider the scattering problems using approximate methods. Among such methods, it is worth mentioning the eikonal approximation of quantum electrodynamics  \cite{AIA96, Glauber}. This approximation has a wider scope of application compared to the frequently used Born approximation, and takes into account the quantum nature of the scattering particles, unlike the classical approach. In this work, within the framework of the eikonal approximation, the problem of fast charged particles scattering on a system of parallel planes with a uniform distribution of atoms in each of them is considered. Uniform distribution of atoms in the plane is a model to simplify consideration of scattering on atomic planes in crystal. Also, we use the idea of continuous potential \cite{Lind} to analytically obtain the differential cross section for this problem, which significantly simplifies the study of the scattering process in this case. This work is a continuation of the study on the scattering of fast charged particles by a single atomic plane in the quasiclassical approximation \cite{Eik}, as well as the study on rainbow scattering of such particles by a single atomic plane \cite{RainC}.

\section{The differential cross section for scattering on isolated structures}
\begin{figure}[h]
      \centering
	   \includegraphics[width=0.5\textwidth]{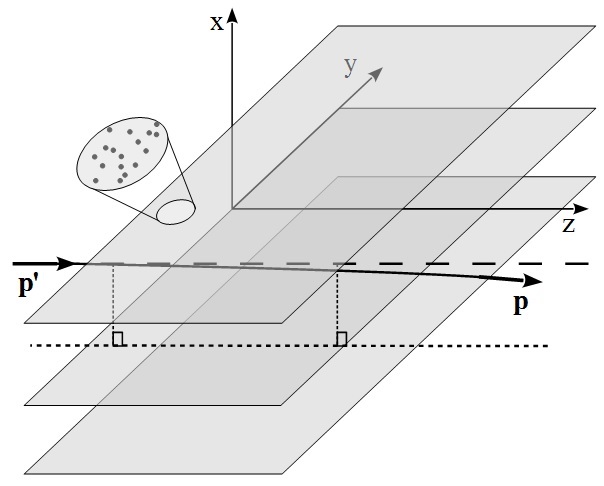}
	\caption{The scattering on the set of parallel planes with uniformly distributed atoms in each plane}
 \label{fig_np}
\end{figure}

Let us consider the scattering of a fast charged particle with an initial momentum parallel to the $z$-axis on a target consisting of a set of parallel atomic planes (parallel to $(z,y)$-plane) arranged along the $x$-axis such that each of them has a position along the $x$-axis: $x_j=aj$, where $a$ is the interplanar distance, $j$ is an integer number (Fig. \ref{fig_np}). In this case, we will assume that the atoms in each plane are of one element and distributed uniformly. In the eikonal approximation, the distribution of atoms along $z$-axis does not influence the scattering description. So, we need to specify only the distribution law for $y$-component of atoms positions. Let us denote by $\vec{\rho}_n=(x_n,y_n)$ the position of the $n$-th atom in the target. Due to uniform arrangment of atoms in the target, the distribution law for each $y_n$ is 
\begin{eqnarray}\label{eq3_1}
g(y_{n})= \begin{cases}\frac{1}{L_y}, \ -\frac{L_y}{2}\leq y_n \leq \frac{L_y}{2};\\
0, \ |y_n|>\frac{L_y}{2};
\end{cases}
\end{eqnarray}
where $L_y$ is the size of the atomic planes in the $y$-axis direction, the index $n$ takes natural values from 1 to $N$, $N$ is the number of all atoms in the target. We also denote by $L_z$ the length of the target in the direction of particle incidence, $N_x$ is the number of planes in the target, $N_p$ is the number of atoms in each plane (assuming that each plane has the same amount of atoms, $N=N_x N_p$), $n_{yz}$ is the 2D-concentration of atoms in each plane: $N_p=n_{yz}L_y L_z$. For simplicity, we will neglect thermal vibrations of atoms in the plane. We will perform the calculations in the system of units, where the speed of light in vacuum is $c=1$ and the reduced Planck constant is $\hslash=1$.

The scattering amplitude in the eikonal approximation is given by the following formula \cite{AIA96}
\begin{eqnarray}\label{eq3_2}
\acute{a}(\vec{q}_{\perp})=\frac{i}{2\pi} \int\limits_{-\infty}^{\infty} dx \int\limits_{-\infty}^{\infty} dy e^{i \vec{q}_{\perp} \vec{\rho}} \left\{ 1- \exp\left[ i \chi^{(N)}_0 (\vec{\rho}) \right] \right\},
\end{eqnarray}
where $\vec{\rho}=(x,y)$, $\vec{q}_{\perp}=(q_x,q_y)$ is the momentum transfer along the corresponding axes, \linebreak $\chi^{(N)}_0=-Qe\int_{-\infty}^{\infty} dz U^{(N)}(\vec{\rho},z)$, $Qe$ is the charge of the incident particle, $U^{(N)}(\vec{\rho},z)$ is the total potential of the target: $U^{(N)}(\vec{\rho},z)=\sum_{n=1}^N u(\vec{\rho}-\vec{\rho}_n,z)$, $u(\vec{\rho}-\vec{\rho}_n,z)$ is the potential of the individual $n$-th atom. Then
\begin{eqnarray}\label{eq3_3}
\chi^{(N)}_0 (\vec{\rho})=\sum_{n=1}^N \chi_0 (\vec{\rho}-\vec{\rho}_n),
\end{eqnarray}
where $\chi_0 (\vec{\rho}-\vec{\rho}_n)=-Qe\int_{-\infty}^{\infty} dz u(\vec{\rho}-\vec{\rho}_n,z)$.
The differential scattering cross section for $\vec{q}_{\perp} \neq 0$ can be written, as in \cite{Eik}, in the general form
\begin{eqnarray}\label{eq3_4}
\frac{d\sigma}{d^2q_{\perp}}=\frac{1}{4\pi^2} \int\limits_{\mathbb{R}^4}d^2 \rho d^2\rho' e^{i\vec{q}_{\perp}(\vec{\rho}-\vec{\rho}')}e^{i[\chi^{(N)}_0(\vec{\rho})-\chi^{(N)}_0(\vec{\rho}')]},
\end{eqnarray}
where $\int_{\mathbb{R}^k}d^2 \rho d^2\rho'...$ denotes the integral over the entire real $k$-dimensional volume.

Since the exact form of the function $\chi^{(N)}_0$ is too complicated, we average the differential scattering cross section over the positions of the atoms in the planes. Averaging the cross section \eqref{eq3_4} reduces to averaging the quantity $e^{i[\chi^{(N)}_0(\vec{\rho})-\chi^{(N)}_0(\vec{\rho}')]}$:
\begin{eqnarray}\label{eq3_5}
\langle e^{i[\chi^{(N)}_0(\vec{\rho})-\chi^{(N)}_0(\vec{\rho}')]} \rangle = \left(\prod_{n=1}^N \int\limits_{-\infty}^{\infty}dy_n g(y_n) \right)  \times \nonumber \\ 
\times \prod_{j=1}^{N_x} \prod_{k=1}^{N_p} \exp \left(i[\chi_0(x-aj,y-y_n)]- \chi_0(x'-aj,y'-y_n)] \right).
\end{eqnarray}
Since all atoms are distributed according to the same law, the expression \eqref{eq3_5} can be represented as follows
\begin{eqnarray}\label{eq3_6}
\langle e^{i[\chi^{(N)}_0(\vec{\rho})-\chi^{(N)}_0(\vec{\rho}')]} \rangle =  \prod_{j=1}^{N_x} \left\{ \int\limits_{-\infty}^{\infty}dy_0 g(y_0)   \exp \left(i[\chi_0(x-aj,y-y_0)] - \chi_0(x'-aj,y'-y_0)] \right) \right\}^{N_p}.
\end{eqnarray}
Based on \cite{Glauber,Eik}, we write \eqref{eq3_6} as
\begin{eqnarray}\label{eq3_7}
\langle e^{i[\chi^{(N)}_0(\vec{\rho})-\chi^{(N)}_0(\vec{\rho}')]} \rangle = \prod_{j=1}^{N_x} \exp \left\{N_p \left[i \langle \chi_{(j)} - \chi'_{(j)} \rangle - \right. \right. \nonumber \\ 
\left. \left.  -\frac{1}{2} \langle \left( \chi_{(j)}-\chi'_{(j)} \right)^2 \rangle + \frac{1}{2}\langle \chi_{(j)}-\chi'_{(j)} \rangle ^2 + ... \right] \right\},
\end{eqnarray}
which is equivalent to
\begin{eqnarray}\label{eq3_8}
\langle e^{i[\chi^{(N)}_0(\vec{\rho})-\chi^{(N)}_0(\vec{\rho}')]} \rangle =  \exp \left\{N_p \left[i \sum_{j=1}^{N_x} \langle \chi_{(j)}-\chi'_{(j)} \rangle - \right. \right.  \nonumber \\  
\left. \left. -\frac{1}{2} \sum_{j=1}^{N_x} \langle \left( \chi_{(j)}-\chi'_{(j)} \right)^2 \rangle + \frac{1}{2} \sum_{j=1}^{N_x}\langle \chi_{(j)}-\chi'_{(j)} \rangle ^2 + ... \right]\right\},
\end{eqnarray}
where $\chi_{(j)}=\chi_0(x-aj,y-y_0)$, $\chi'_{(j)}=\chi_0(x'-aj,y'-y_0)$, the brackets $\langle ... \rangle$ denote averaging, e.g. \linebreak $\langle \chi_{(j)} \rangle=\int_{-\infty}^{\infty}dy_0 g(y_0) \chi_0(x-aj,y-y_0)$. 

Up to terms linear in the function $\chi_0(\vec{\rho})$, the differential scattering cross section is
\begin{eqnarray}\label{eq3_9}
\frac{d\sigma}{d^2q_{\perp}}=\frac{1}{4\pi^2} \Big|\int\limits_{\mathbb{R}^2} d^2 \rho  e^{i\vec{q}_{\perp}\vec{\rho}} \exp \left[i \sum_{j=1}^{N_x} \langle \chi_{(j)} \rangle \right] \Big|^2,
\end{eqnarray}
which can be expressed as
\begin{eqnarray}\label{eq3_10}
\frac{d\sigma}{d^2q_{\perp}}=\frac{1}{4\pi^2} |\tilde{a}(\vec{q}_{\perp})|^2,
\end{eqnarray}
where, for convenience, we introduce the "renormalized" scattering amplitude $\tilde{a}(\vec{q}_{\perp})$: 
\begin{eqnarray}\label{eq3_11}
\tilde{a}(\vec{q}_{\perp})=\int\limits_{\mathbb{R}^2} d^2 \rho e^{i \vec{q}_{\perp} \vec{\rho}} \left\{ 1- \exp\left[ i \sum_{j=1}^{N_x} \langle \chi_0(\vec{\rho}-\vec{\rho}_j) \rangle \right] \right\},
\end{eqnarray}
remembering that $\int_{\mathbb{R}^2} d^2 \rho e^{i \vec{q}_{\perp} \vec{\rho}}=0$ for $\vec{q}_{\perp} \neq 0$ since $\int_{\mathbb{R}^2} d^2 \rho e^{i \vec{q}_{\perp} \vec{\rho}}=4\pi^2 \delta(\vec{q}_\perp)$. So, our considerations led us to the differential scattering cross section in the approximation of continuous potential \cite{Lind}.

One may consider the formula \eqref{eq3_11} from the following point of view. Let the total potential of the target consist of the potentials of $M$ "substructures" (e.g. atoms, atomic strings, atomic planes, etc.). Then, if $\chi_0$ is a function for such a substructure and $\vec{\rho}_k$ is the position of such a substructure (assuming that substructures are parallel to $(x,y)$ plane), then
\begin{eqnarray}\label{eq3_13}
\tilde{a}(\vec{q}_{\perp})=\int\limits_{\mathbb{R}^2} d^2 \rho e^{i \vec{q}_{\perp} \vec{\rho}} \left\{ 1- \exp\left[ i \sum_{k=1}^M \chi_0(\vec{\rho}-\vec{\rho}_k) \right] \right\}.
\end{eqnarray}
If the characteristic distance at which the potential fades is $\sim R$ (the screening radius), and the characteristic distance between the "substructures" is $\sim a: R \ll a$, then
\begin{eqnarray}\label{eq3_14}
\tilde{a}(\vec{q}_{\perp}) \approx \int\limits_{-\infty}^{l_1} dx \int\limits_{-\infty}^{s_1} dy e^{i \vec{q}_{\perp} \vec{\rho}} \left\{ 1- \exp\left[ i \chi_0(\vec{\rho}-\vec{\rho}_1) \right] \right\} + \nonumber \\  
+ \int\limits_{l_1}^{l_2} dx \int\limits_{s_1}^{s_2} dy e^{i \vec{q}_{\perp} \vec{\rho}} \left\{ 1- \exp\left[ i \chi_0(\vec{\rho}-\vec{\rho}_2) \right] \right\} +  ... +  \nonumber \\ 
+  \int\limits_{l_{m-1}}^{l_m} dx \int\limits_{s_{m-1}}^{s_m} dy e^{i \vec{q}_{\perp} \vec{\rho}} \left\{ 1- \exp\left[ i \chi_0(\vec{\rho}-\vec{\rho}_m) \right] \right\} + ... + \nonumber \\  
+   \int\limits_{l_{M-1}}^{\infty} dx \int\limits_{s_{M-1}}^{\infty} dy e^{i \vec{q}_{\perp} \vec{\rho}} \left\{ 1- \exp\left[ i \chi_0(\vec{\rho}-\vec{\rho}_M) \right] \right\}.
\end{eqnarray}
The integration intervals are chosen so that $\forall m \ x_m \in (l_{m-1},l_m), y_m \in (s_{m-1},s_m): \ |l_m-x_m|,|l_{m-1}-x_m| \gg R, \ |s_m-y_m|,|s_{m-1}-y_m| \gg R$. Due to the rapid decrement of the atomic potential, $\forall m \ \chi_0(l_m-x_m,s_m-y_m) \approx 0$, $\chi_0(x_m-l_{m-1},y_m-s_{m-1}) \approx 0$ so each of the integrals in \eqref{eq3_14} can be approximately written as
\begin{eqnarray}\label{eq3_16}
\tilde{a}(\vec{q}_{\perp}) \approx \int\limits_{-\infty}^{\infty} dx \int\limits_{-\infty}^{\infty} dy e^{i \vec{q}_{\perp} \vec{\rho}} \left\{ 1- \exp\left[ i \chi_0(\vec{\rho}-\vec{\rho}_1) \right] \right\} + \nonumber \\  
+ \int\limits_{-\infty}^{\infty} dx \int\limits_{-\infty}^{\infty} dy e^{i \vec{q}_{\perp} \vec{\rho}} \left\{ 1- \exp\left[ i \chi_0(\vec{\rho}-\vec{\rho}_2) \right] \right\} +  ... + \nonumber \\ 
+   \int\limits_{-\infty}^{\infty} dx \int\limits_{-\infty}^{\infty} dy e^{i \vec{q}_{\perp} \vec{\rho}} \left\{ 1- \exp\left[ i \chi_0(\vec{\rho}-\vec{\rho}_m) \right] \right\} + ... + \nonumber \\  
+  \int\limits_{-\infty}^{\infty} dx \int\limits_{-\infty}^{\infty} dy e^{i \vec{q}_{\perp} \vec{\rho}} \left\{ 1- \exp\left[ i \chi_0(\vec{\rho}-\vec{\rho}_M) \right] \right\}.
\end{eqnarray}
Let us denote one of integrals of the sum as
\begin{eqnarray}\label{eq3_17}
\tilde{a}_m=\int\limits_{\mathbb{R}^2} d^2 \rho e^{i \vec{q}_{\perp} \vec{\rho}} \left\{ 1- \exp\left[ i \chi_0(\vec{\rho}-\vec{\rho}_m) \right] \right\}.
\end{eqnarray}
Substitutions $\xi=x-x_m$, $\eta=y-y_m$ result in
\begin{eqnarray}\label{eq3_18}
\tilde{a}_m=\exp[i(q_x x_m + q_y y_m)]  \int\limits_{-\infty}^{\infty} d\xi \int\limits_{-\infty}^{\infty} d\eta \ \exp[i(q_x \xi + q_y \eta)] \left\{ 1- e^{\left[ i \chi_0(\xi,\eta) \right]} \right\},
\end{eqnarray}
which we write in a form
\begin{eqnarray}\label{eq3_18_1}
\tilde{a}_m=\exp[i(q_x x_m + q_y y_m)] \ \tilde{a}^{(1)}, 
\end{eqnarray}
where $\tilde{a}^{(1)}$ is "renormalized" amplitude for scattering on a single "substructure". So, the scattering amplitude becomes
\begin{eqnarray}\label{eq3_19}
\tilde{a}=\tilde{a}^{(1)} \sum_{k=1}^{M} e^{i\vec{q}_{\perp}\vec{\rho}_k}
\end{eqnarray}
and the differential cross section for scattering on all $M$ substructures is
\begin{eqnarray}\label{eq3_22}
\frac{d\sigma^{(M)}_{eik}}{d^2q_{\perp}}=\frac{d\sigma^{(1)}_{eik}}{d^2q_{\perp}} D_M,
\end{eqnarray}
where $\frac{d\sigma^{(1)}_{eik}}{d^2q_{\perp}}$ is the differential cross section for scattering on a single substructure, $D_M=\Big| \sum_{k=1}^{M} e^{i\vec{q}_{\perp}\vec{\rho}_k} \Big|^2$. $D_M$ is defined up to a certain shift (with respect to the shift of the coordinate system, $D_M$ is invariant). Also, up to this shift, we can define $S_M= \sum_{k=1}^{M} e^{i\vec{q}_{\perp}\vec{\rho}_k}$. Several $S_M$ may correspond to one $D_M$, since $S_M$ depends on the shift of the coordinate system. $D_M$ coincides with the Laue-Bragg interference factor \cite{AIA96}.
\section{The differential cross section for scattering on a system of parallel atomic planes}
Let us consider the case of a fast charged particle scattering on a system of parallel planes with a uniform distribution of atoms in each of them. Let us determine the positions of the planes for an even and odd amount of planes:
\begin{eqnarray}\label{eq3_23}
\{x_k\}_{k=1}^M=\begin{cases} \{ak\}_{k=-(M-1)/2}^{(M-1)/2}, \ M=2m+1; \\
\{a(k-1/2)\}_{k=-M/2+1}^{M/2}, \ M=2m. \end{cases}
\end{eqnarray}
For an even number of planes, let us locate the origin of coordinates along the $x$-axis in the middle so that the planes closest to it were located at the same distance from the origin; for an odd number of planes, let us locate the central plane in the position $x_{(M+1)/2}=0$. So, in any of these cases, the planes are located symmetrically relative to $x=0$. Then
\begin{eqnarray}\label{eq3_24}
S_M= \begin{cases} 1 + 2 \sum\limits_{k=1}^{(M-1)/2} \cos[q_xak], \ M=2m+1; \\
2\sum\limits_{k=1}^{M/2} \cos[q_xa(k-1/2)], \ M=2m. \end{cases}
\end{eqnarray}
Let us assume that the potential of an individual atom is the screened Coulomb potential ($u(\vec{r})=\frac{Ze}{r}\exp(-\frac{r}{R})$, where $Z$ is nuclear charge number of atoms in the target \cite{AIA96}). Then, for a single plane located at $x=0$, the $\chi_0$-function averaged over the positions of atoms in the plane is as follows
\begin{eqnarray}\label{eq3_25}
\chi_0= A  \exp \left[-\frac{|x|}{R} \right],
\end{eqnarray}
where $A=2ZQ \alpha N_{p} \frac{R}{L_y}$, $\alpha$ is  fine-structure constant. However, it is more usefull to define $A=2ZQ \alpha n_{yz}RL_z$ through the 2D-concentration $n_{yz}$: $N_p=n_{yz}L_zL_y$.
 
For the $\chi_0$-function, which depends only variable $x$, the differential scattering cross section becomes
\begin{eqnarray}\label{eq3_26}
\frac{d\sigma^{(1)}}{dq_x}=\frac{L_y}{2\pi} |\tilde{a}^{(1)}({q}_{x})|^2,
\end{eqnarray}
where
\begin{eqnarray}\label{eq3_26_1}
\tilde{a}^{(1)}({q}_{x})=\int_{-\infty}^{\infty} dx e^{i q_x x} \left\{ 1- \exp\left[ i \chi_0(x) \right] \right\},
\end{eqnarray}
since $\Big|\int_{-\infty}^{\infty} dy e^{i q_y y}  \Big|^2=2\pi L_y \delta(q_y)$. The integral in \eqref{eq3_26_1} can be taken analytically for $\chi_0(x)$ defined  by \eqref{eq3_25}:
\begin{eqnarray}\label{eq3_26_2}
\tilde{a}^{(1)}=-\frac{A}{q_x R} \left\{(-iA)^{-1+i q_x R}\gamma(1-iq_x R, -iA) - (-iA)^{-1-i q_x R}\gamma(1+iq_x R, -iA) \right\},
\end{eqnarray}
where $\gamma(s, x)$ is the lower incomplete gamma function.
It is easier to compute \eqref{eq3_26_2} in the form of the following series (the following expression is exact)
\begin{eqnarray}\label{eq3_27}
\tilde{a}^{(1)}=-2 \sum_{k=0}^{\infty} \frac{(iA)^{k+1}}{k! \left[ (k+1)^2 + q_x^2 R^2 \right]}.
\end{eqnarray}
\section{Discussion}
\begin{figure}[!h]
 \centering
\includegraphics[width=0.5\textwidth]{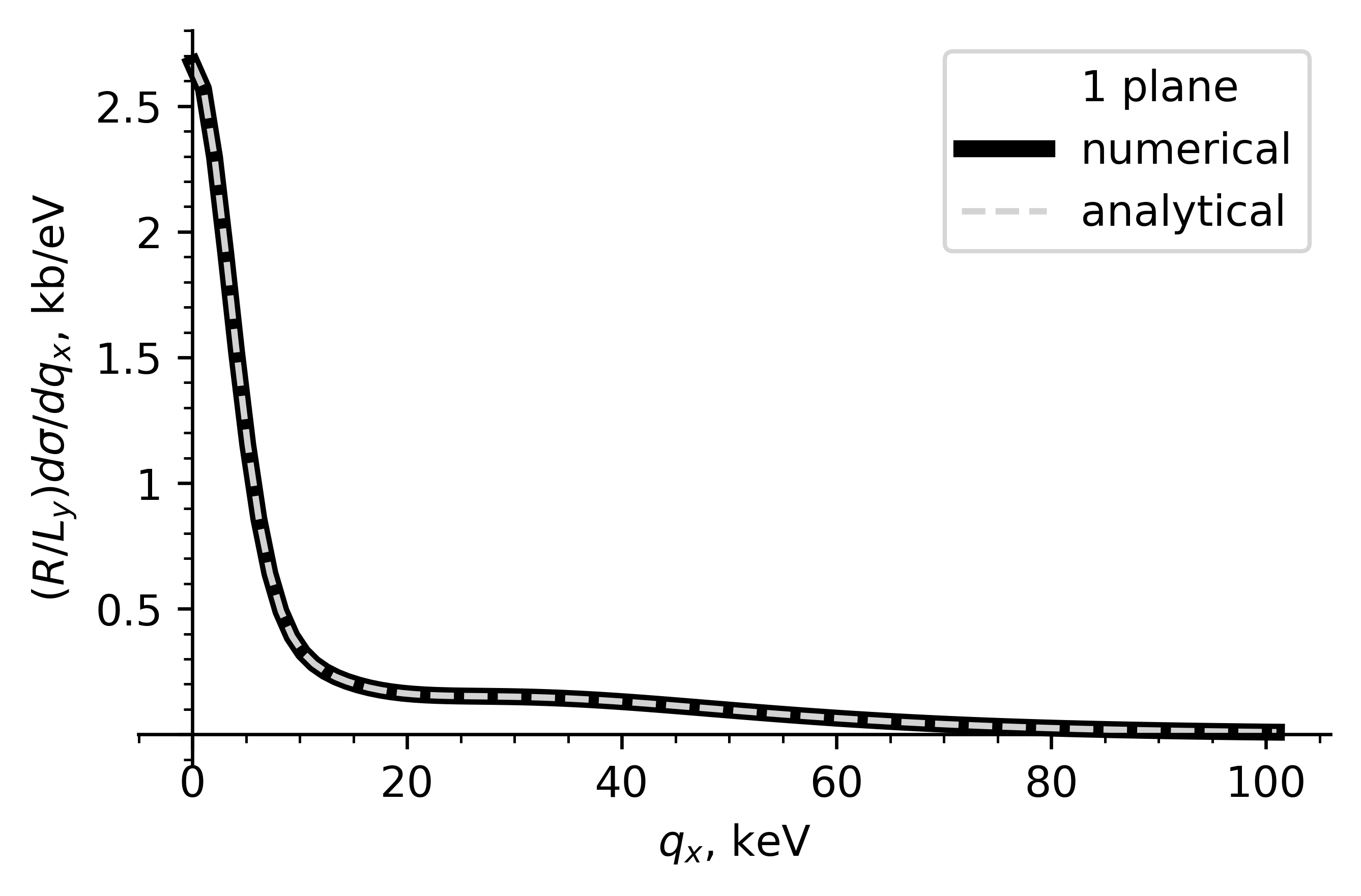}
\caption{The differential cross section of fast charged particles scattering on a single atomic plane, obtained numerically and analytically}
 \label{fig_pl1}
\end{figure}

Based on the considerations above, the differential cross sections of fast charged particles scattering on 1-5 (110)-planes of silicon atoms at $A=10$ were calculated. Fig. \ref{fig_pl1} shows the differential scattering cross section for one plane, calculated using the analytically determined scattering amplitude \eqref{eq3_27} and using the numerically integrated expression \eqref{eq3_26_1}. Rainbow scattering in the case of scattering on a single plane is not observed here, in contrast to the results of \cite{RainC}, because thermal vibrations of the target atoms, which lead to the rainbow scattering, are not taken into account here. The inclusion of thermal vibrations in this problem is a subject for separate consideration. Without accounting for thermal vibrations, the differential scattering cross section reflects only the general features of the scattering.

\begin{figure}[!h]
      \centering
	   \begin{subfigure}{0.49\linewidth}
		\includegraphics[width=\textwidth]{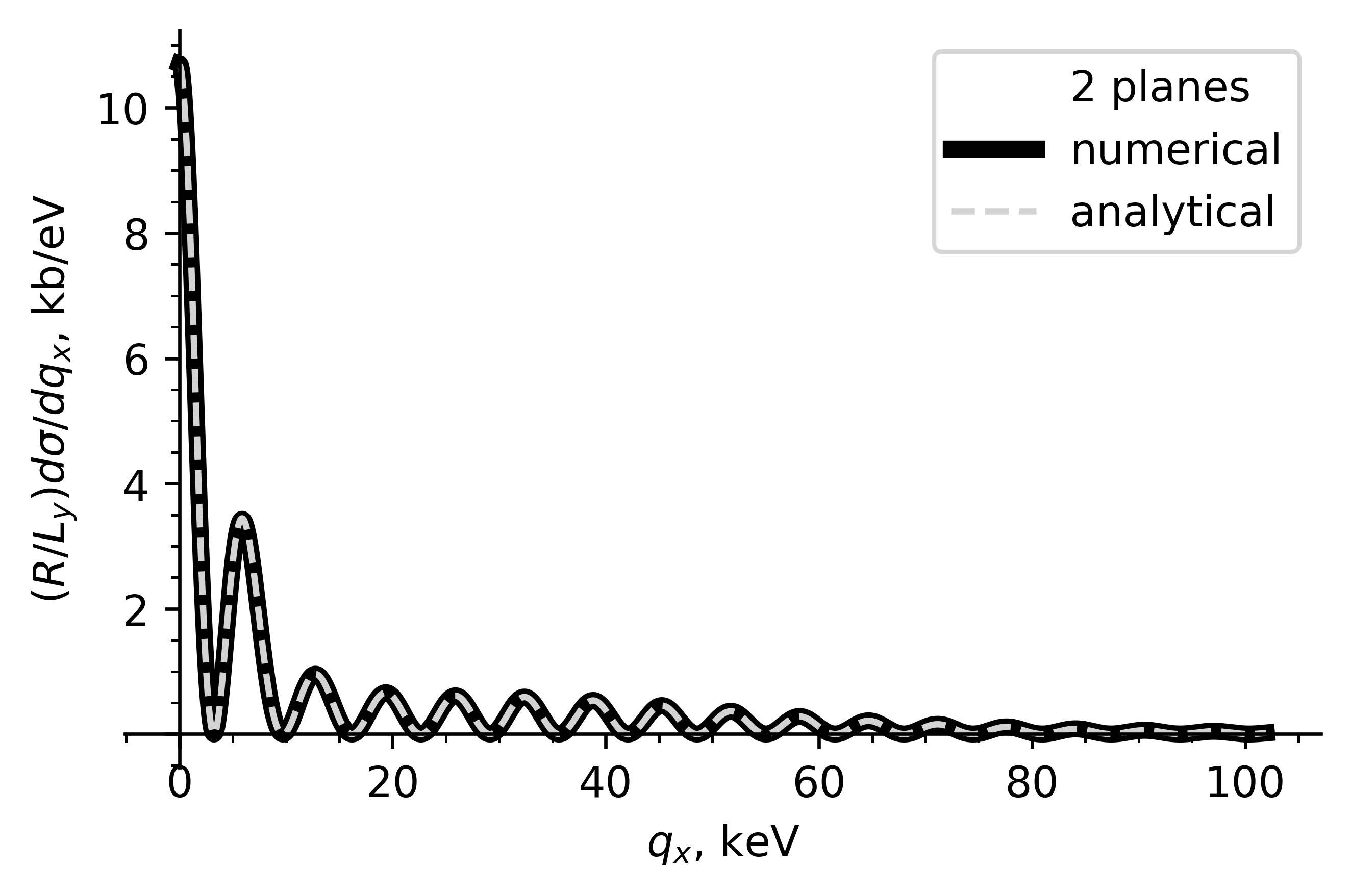}
		\caption{$q_x \in [0, 102]$ keV}
		\label{fig:subfig4.2_1}
	   \end{subfigure}
	     \begin{subfigure}{0.49\linewidth}
		 \includegraphics[width=\textwidth]{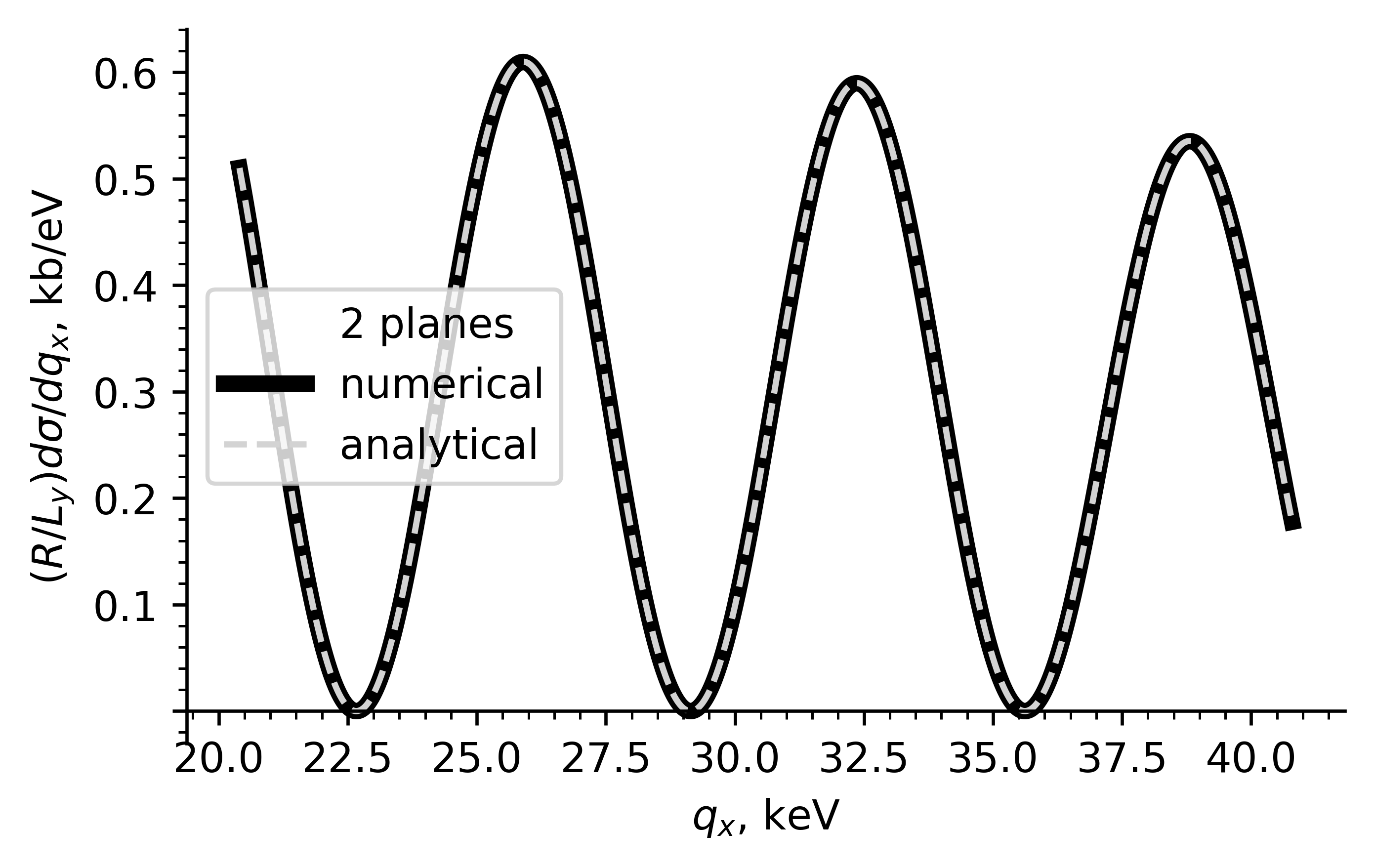}
		\caption{$q_x \in [20.4, 40.8]$ keV}
		 \label{fig:subfig4.2_2}
	      \end{subfigure}
\caption{The same as in Fig. \ref{fig_pl1} for 2 planes}
 \label{fig_pl2}
\end{figure}

\begin{figure}[ht]
      \centering
\begin{subfigure}{\linewidth}
\centering
		\includegraphics[width=0.5\textwidth]{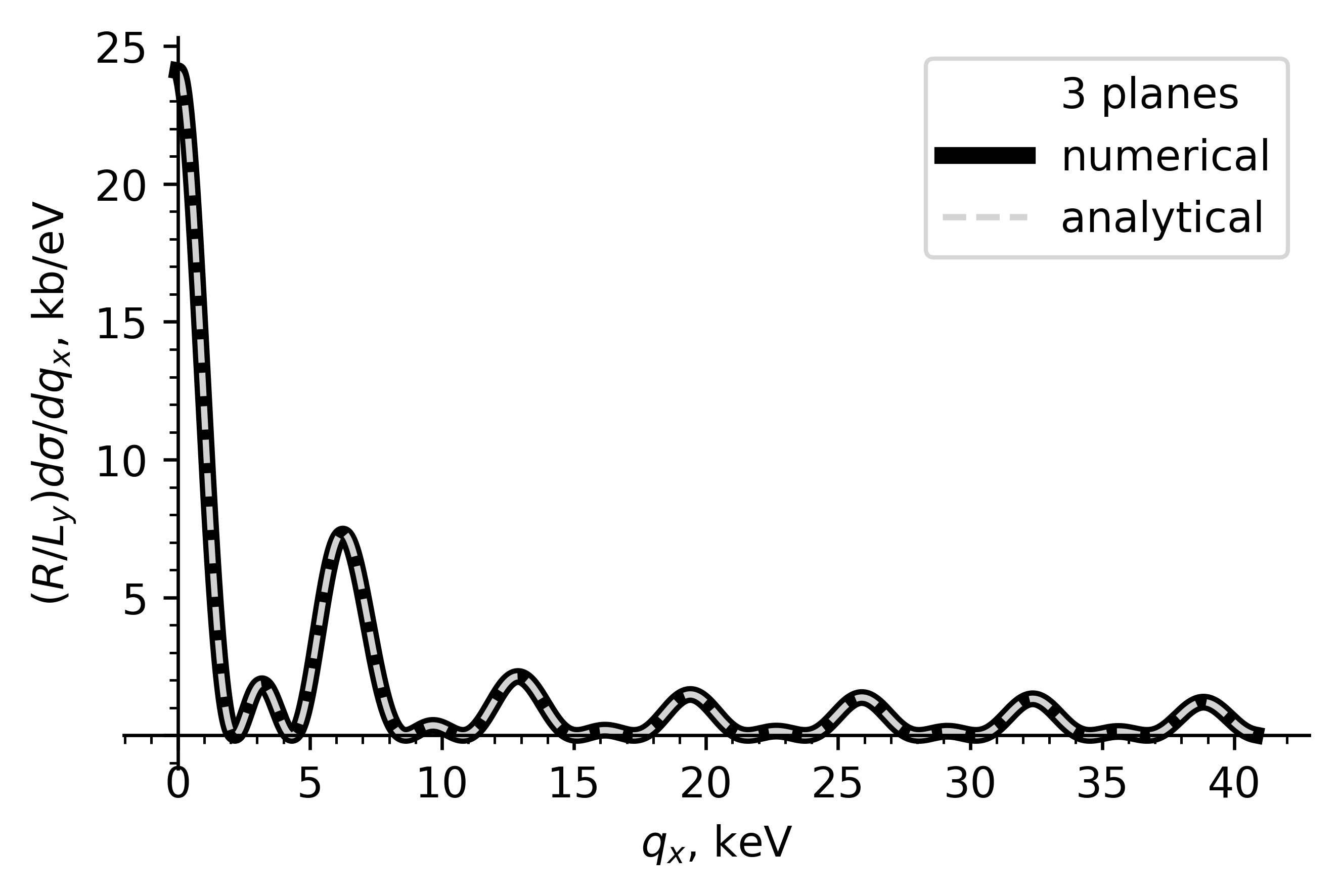}
		\caption{$q_x \in [0, 40.8]$ keV}
		\label{fig:subfig4.2_3}
	    \end{subfigure}
\vfill
	   \begin{subfigure}{0.49\linewidth}
		\includegraphics[width=\textwidth]{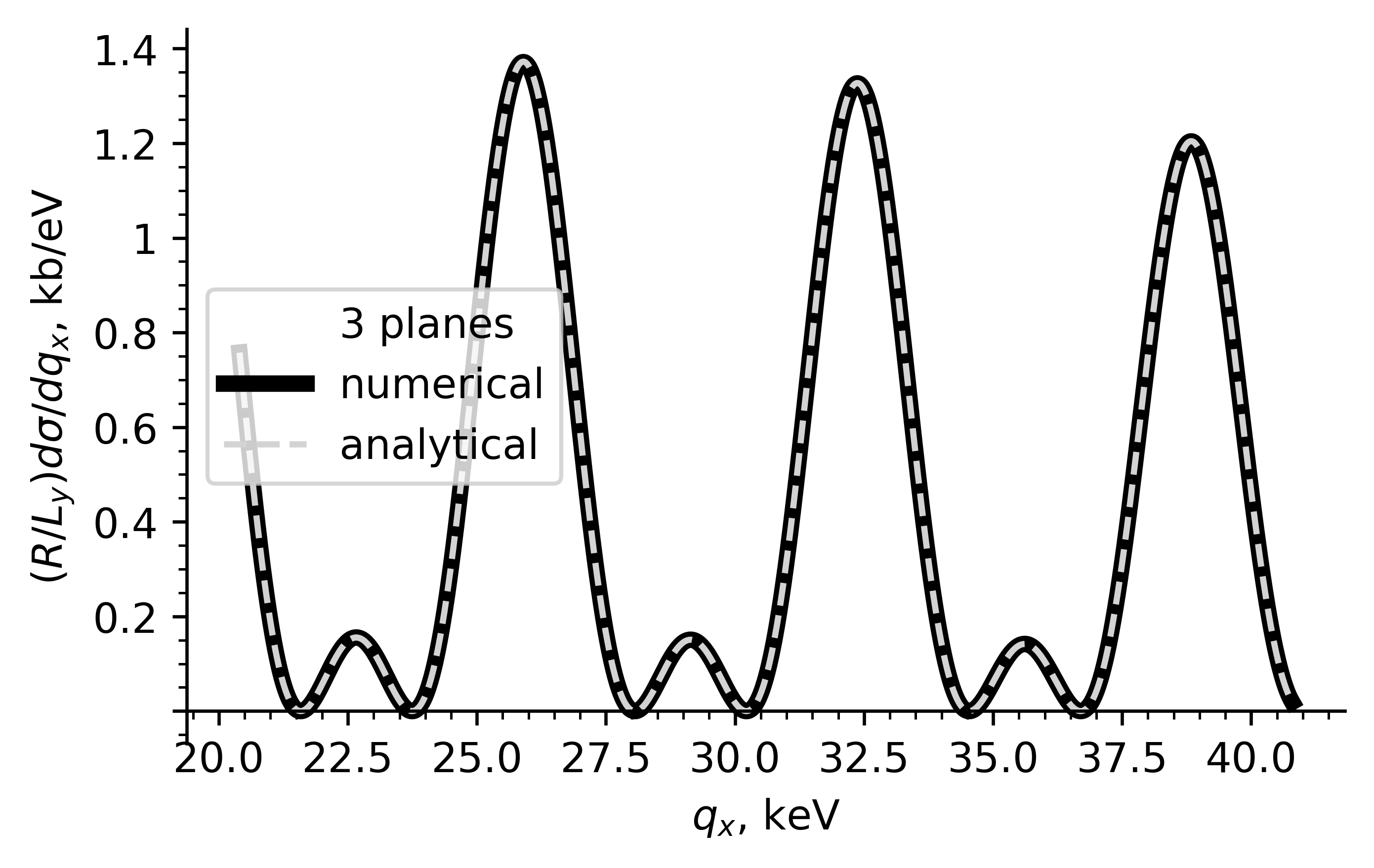}
		\caption{$q_x \in [20.4, 40.8]$ keV}
		\label{fig:subfig4.2_1}
	   \end{subfigure}
	     \begin{subfigure}{0.49\linewidth}
		 \includegraphics[width=\textwidth]{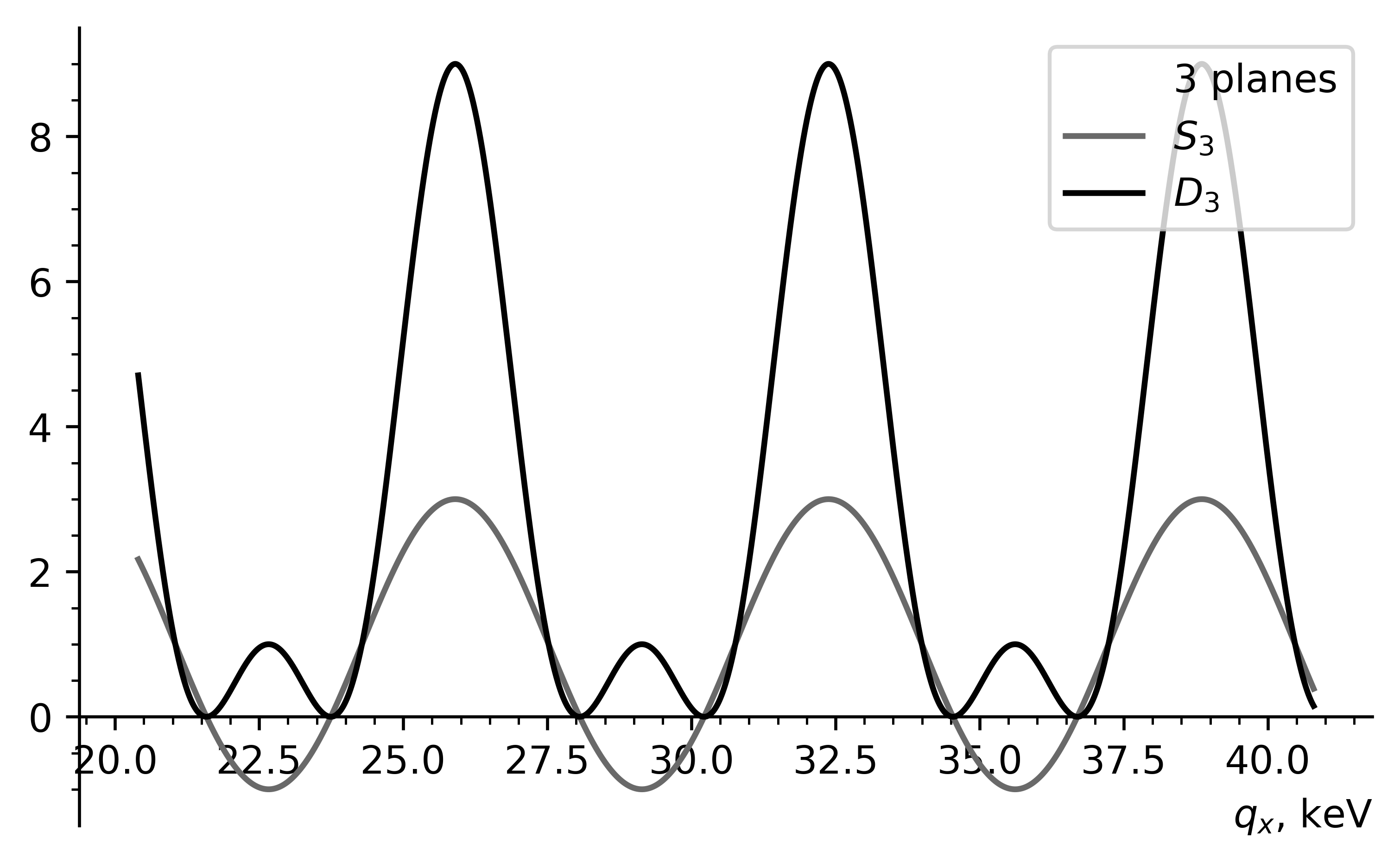}
		 \caption{$S_3, \ D_3$}
		 \label{fig:subfig4.2_2}
	      \end{subfigure}
\caption{The same as in Fig. \ref{fig_pl1} for 3 planes and $S_3$, $D_3$}
 \label{fig_pl3}
\end{figure}

\begin{figure}[!ht]
      \centering
\begin{subfigure}{\linewidth}
      \centering
		\includegraphics[width=0.5\textwidth]{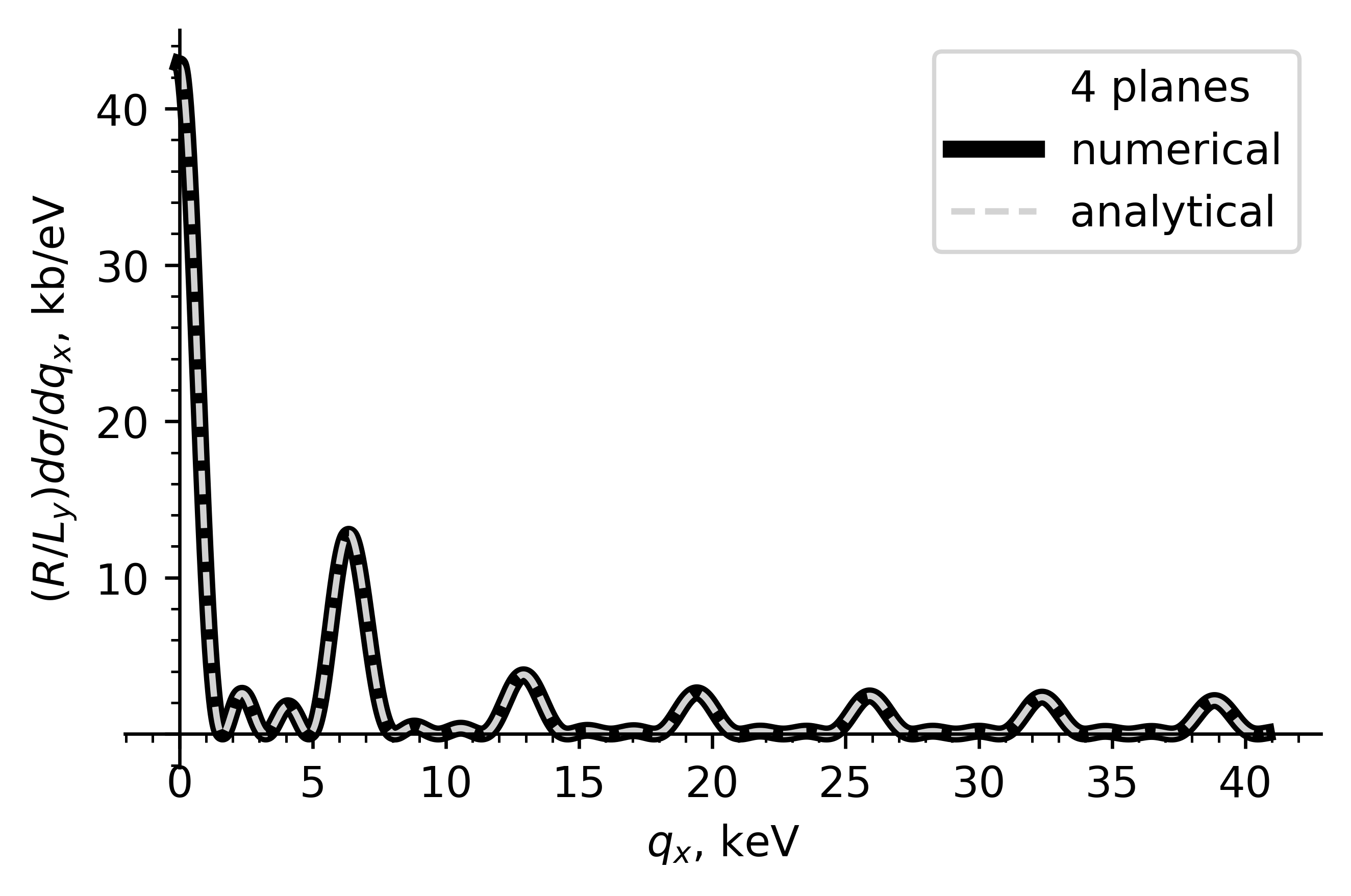}
		\caption{$q_x \in [0, 40.8]$ keV}
		\label{fig:subfig4.2_3}
	    \end{subfigure}
\vfill
	   \begin{subfigure}{0.49\linewidth}
		\includegraphics[width=\textwidth]{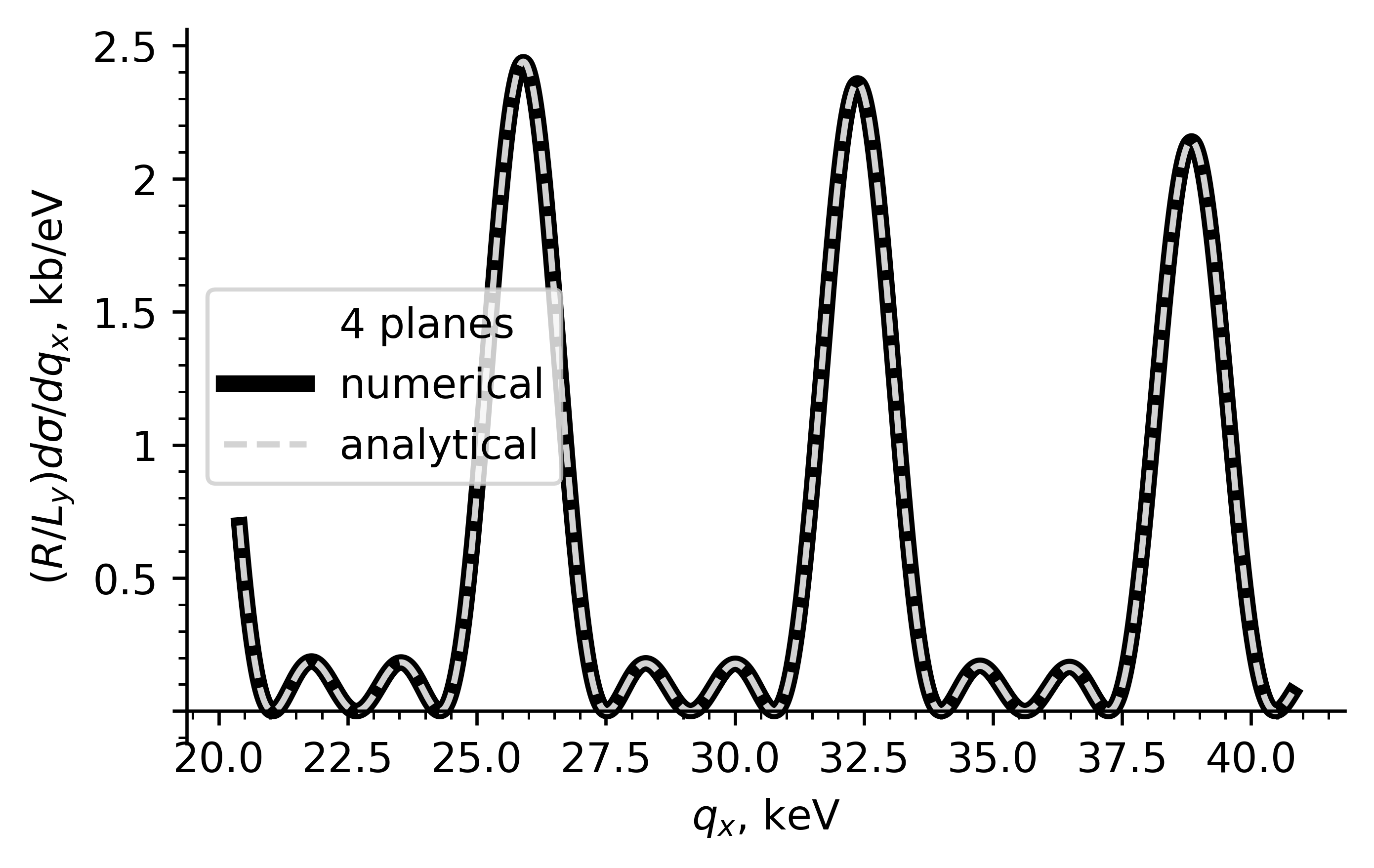}
		\caption{$q_x \in [20.4, 40.8]$ keV}
		\label{fig:subfig4.2_1}
	   \end{subfigure}
	     \begin{subfigure}{0.49\linewidth}
		 \includegraphics[width=\textwidth]{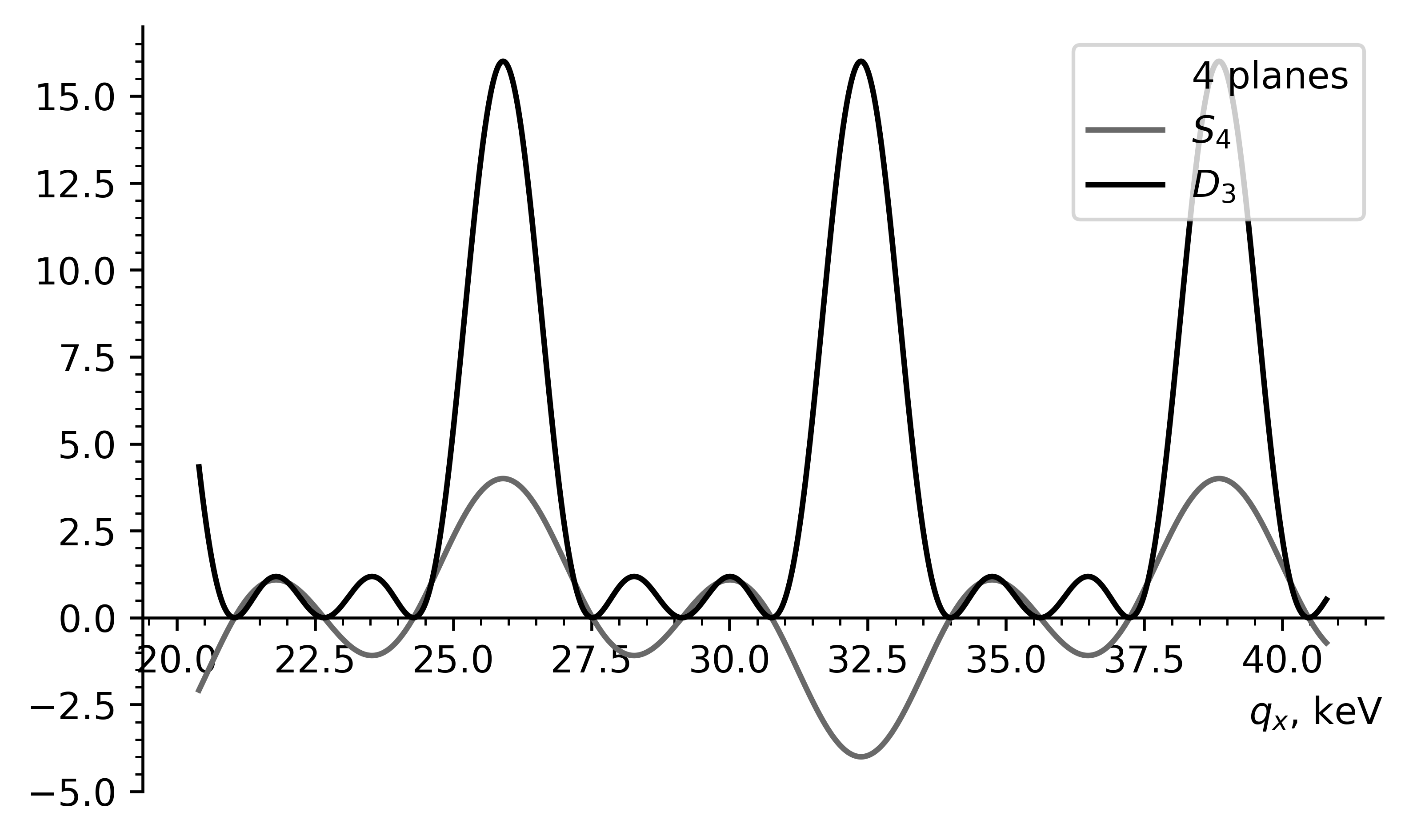}
		  \caption{$S_4, \ D_4$}
		 \label{fig:subfig4.2_2}
	      \end{subfigure}
\caption{The same as in Fig. \ref{fig_pl1} for 4 planes and $S_4$, $D_4$}
 \label{fig_pl4}
\end{figure}

Figs. \ref{fig_pl2}-\ref{fig_pl4} show the differential scattering cross sections on 2, 3, and 4 planes, respectively, obtained both analytically by substituting \eqref{eq3_26} with the analytically calculated scattering amplitude \eqref{eq3_27} into the formula \eqref{eq3_22}, and numerically by numerically integrating the expression \eqref{eq3_13} over the variable $x$ and substituting it to \eqref{eq3_10}. Figs. \ref{fig_pl1}-\ref{fig_pl4} indicate that the numerical and analytical results agree well. However, obtaining differential cross sections analytically requires much less computational resources and time compared to the method based on numerical integration. It is also illustrated in Figs. \ref{fig_pl2}-\ref{fig_pl4} that it is the form of $D_M$ that determines the oscillation patterns of the corresponding differential cross sections (noting that $D_2=4\cos^2[q_xa/2]$).

\begin{figure}[!ht]
 \centering
\includegraphics[width=0.5\textwidth]{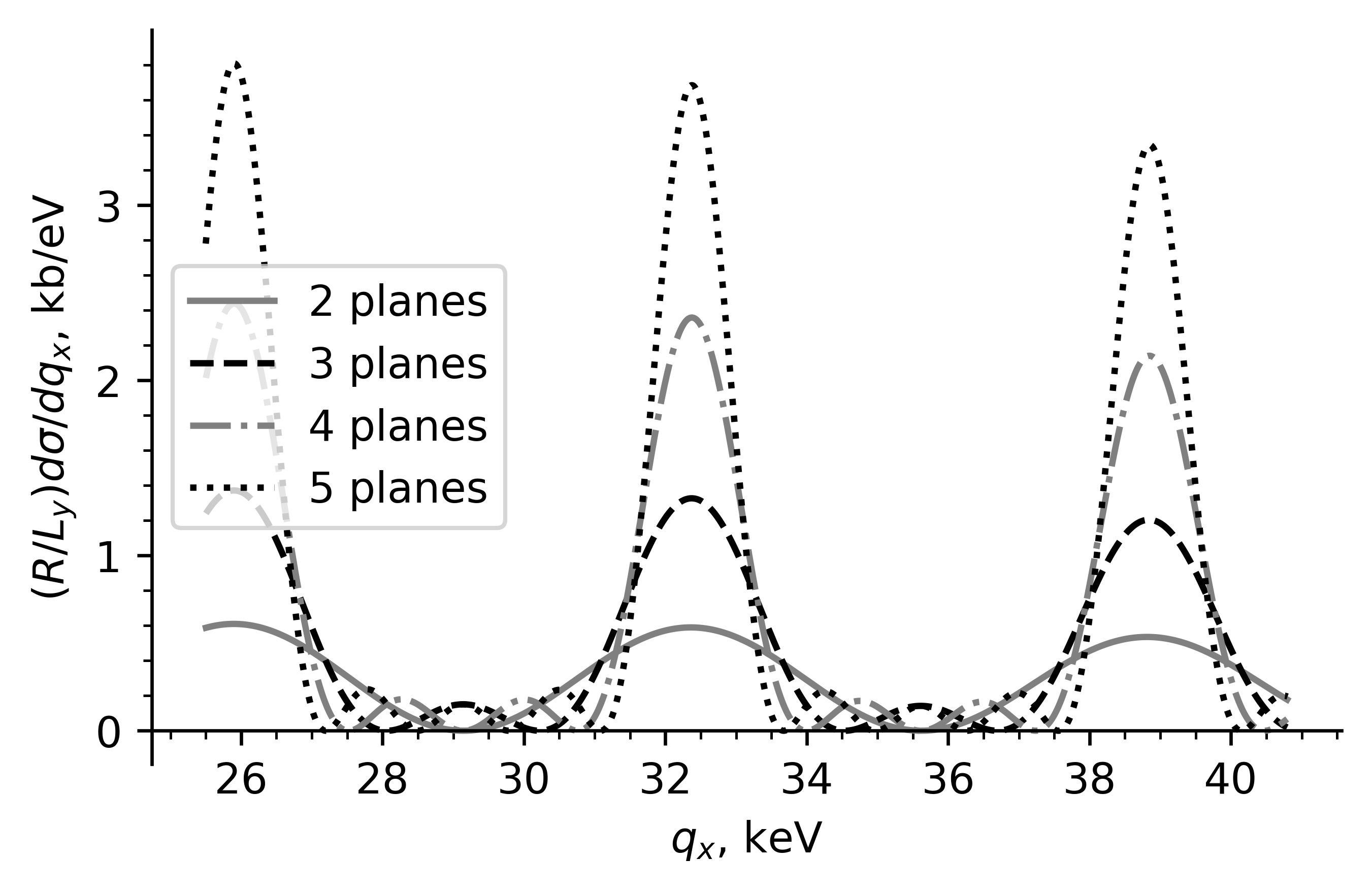}
\caption{Differential cross sections of fast charged particles scattering on 2-5 atomic planes at $q_x \in [25.5, 40.8]$ keV}
 \label{fig_pl2-5}
\end{figure}

Comparing the differential scattering cross sections for 2-5 planes  (Fig. \ref{fig_pl2-5}),  one may notice the sensitivity of the differential cross sections to the number of atomic planes in the target. This feature can be used for target diagnostics.

We note that the largest peaks in Fig. \ref{fig_pl2-5} for all numbers of planes have the same position. For a very large number of planes $N_x \gg 1$ in the target, $D_{N_x} \sim \sum_g \delta(q_x-g(j))$, where $g(j)=\frac{2\pi j}{a}$ are the reciprocal lattice vectors and $j=0, \pm 1, \pm 2, ...$. The differential scattering cross section, in this case:  
\begin{eqnarray}\label{eq3_28}
\frac{d\sigma}{dq_x}=\frac{L_y N_x}{a} |\tilde{a}^{(1)}|^2 \sum_{j=-\infty}^{\infty} \delta\left(q_x-g(j) \right)
\end{eqnarray}
has common features with the coherent differential cross section for the problem of scattering of fast charged particles on a system of parallel planes with a uniform distribution of atoms considered in the Born approximation \cite{Born}.

Also, it is important to define the applicability region of these calculations. Based on \cite{RainC}, the eikonal approximation for this case is applicable  when $pR/(2Z|Q|\alpha) \gg n_{yz} L_z^2$. For chosen particles and other fixed target parameters, $A$ defines longitudinal size of the target. For example, for incident electrons or positrons (|Q|=1) and for parameters of (110) Si planes, $L_z[\mu \text{m}]=0.026 \ A$, meaning we considered the target of 0.26 $\mu \text{m}$ longitudinal size. The necessary energy of the incident electrons (positrons) for the applicability condition is $\varepsilon[\text{GeV}] \gg 20 \ (L_z[\mu \text{m}])^2$, meaning $\varepsilon \gg  1.35 \ \text{GeV}$ for this case. For thicker targets, the required energy of incident particles increases quadratically.



\FloatBarrier
\begingroup
\section*{Acknowledgements}
{Author thanks God and His Blessed Mother for saving us and our Ukraine. The work was partially supported by the National Academy of Sciences of Ukraine (project 0124U002155). The author is deeply thankful to her late research supervisor, an academic of NAS of Ukraine N.F. Shul'ga, who introduced this problem to her and gave a lot of necessary knowledge. Author also acknowledges the fruitful discussions with S.P. Fomin, I.V. Kyryllin, S.V. Trofymenko.  
\par}

\endgroup

\end{document}